\pdfoutput=1

\documentclass[aps,prl,amsmath,amssymb,floatfix,twocolumn,amsmath,superscriptaddress,twocolumn,nofootinbib,tighten,letterpaper]{revtex4-2}

\usepackage[utf8]{inputenc}
\usepackage{bm}
\usepackage{siunitx}
\usepackage{enumerate}
\usepackage{amsfonts}
\usepackage{amsmath}
\usepackage{amssymb}
\usepackage{color}
\usepackage{soul}
\usepackage{float}
\usepackage{graphicx}
\usepackage{graphics}
\usepackage{physics}
\usepackage{verbatim}
\usepackage[colorlinks,allcolors=blue]{hyperref}
\usepackage{tikz-feynman}
\usepackage{xcolor,soul}
\usepackage{cancel}
\usepackage{tikz}
\usetikzlibrary{angles, quotes, intersections}
\usepackage[capitalise]{cleveref}

\newcommand{\bs}[1]{\boldsymbol{#1}}
\let\propto=\sim
\let\epsilon=\varepsilon

\renewcommand{\vec}[1]{\bm{#1}}

\definecolor{DarkRed}{rgb}{0.80,0,0}
\definecolor{DarkGray}{rgb}{0.7,0.7,0.7}

\newcommand{\prlsection}[1]{\textit{#1}.\kern0.05em---\kern0.05em\ignorespaces}
\begin{document}

\title{Anisotropic marginal Fermi liquid for Coulomb interacting generalized Weyl fermions}

\author{Gabriel Malav\'e}
\affiliation{Department of Physics and Astronomy, Rutgers University, Piscataway, NJ 08854, USA}
\affiliation{Facultad de Física, Pontificia Universidad Católica de Chile, Vicuña Mackenna 4860, Santiago 8331150, Chile}

\author{Rodrigo Soto-Garrido}
\affiliation{Facultad de Física, Pontificia Universidad Católica de Chile, Vicuña Mackenna 4860, Santiago 8331150, Chile}

\author{Bitan Roy}
\affiliation{Department of Physics, Lehigh University, Bethlehem, Pennsylvania, 18015, USA}
\affiliation{Centre for Condensed Matter Theory, Department of Physics, Indian Institute of Science, Bengaluru 560012, India}

\author{Vladimir Juri\v{c}i\'c}
\affiliation{Departamento de F\'isica, Universidad T\'ecnica Federico Santa Mar\'ia, Casilla 110, Valpara\'iso, Chile}

\begin{abstract}
Owing to a power-law anisotropy in the quasiparticle dispersion, yielding an enhanced density of states, the effects of long-range Coulomb interaction get amplified in three-dimensional generalized Weyl semimetals, characterized by integer monopole charge $n>1$ of the underlying Weyl nodes. Using a Wilsonian renormalization group approach controlled by a large-$N$ expansion with $N$ as the number of Weyl fermion flavors and a gauge-consistent regularization fixed by the Ward-Takahashi identity, we uncover for $n\ge 2$ an extended interaction-dominated scaling regime with intrinsically anisotropic dynamic Coulomb screening, a finite fermionic anomalous dimension, and a power-law suppression of the quasiparticle residue, yielding an \emph{anisotropic} marginal non-Fermi liquid at intermediate energies. Ultimately, the effective fine structure constant flows to zero, albeit only logarithmically slowly, so the marginal Fermi liquid phenomenology emerges as a broad crossover, controlled by a slowly running coupling. By contrast, for $n=1$ the system retains an isotropic marginal Weyl-liquid character. These predictions can be tested via scaling in thermodynamics (specific heat and compressibility), direction-dependent optical conductivity, and by anisotropic broadening of the single-particle spectral function in angle-resolved photoemission spectroscopy.

\end{abstract}
\maketitle


\emph{Introduction}. Weyl semimetals (WSMs) realize a gapless topological phase of quantum matter in which conduction and valence bands touch at isolated points in the Brillouin zone, known as the Weyl nodes that act as monopole sources and sinks of Berry curvature in momentum space. WSMs have become a central platform for exploring topology in metals and semimetals, motivated by their distinctive bulk and surface signatures and a rapidly expanding set of material realizations \cite{Armitage2018RMP, Chiu2016RMP, Hasan2017AnnRevCondMattPhys, Burkov2018WeylMetals}. Their hallmark signatures include surface Fermi arcs~\cite{Xu2015ScienceTaAs, Lv2015PRXTaAs, Lv2015NatPhysTaAs}, topological electromagnetic responses governed by Abelian and non-Abelian chiral anomalies~\cite{ZyuzinBurkov2012TopologicalResponse, SonSpivak2013NegativeMR, Grushin2012, GoswamiTewari2013, VazifehFranz2013, Dantas2020}, and a disorder-driven non-Anderson quantum phase transition into a diffusive metal~\cite{Fradkin1986, Koshino2014, Goswami2011, Herbut2014, Brouwer2014, Syzranov2015, Pixley2015, Bera2016, Slager2018, Fedorenko2018}. Our study is applicable to Weyl systems, where the chemical potential resides at or close to the Weyl nodes, a condition typically satisfied in 227 pyrochlore iridates with multipolar magnetic orders~\cite{Wan2011PyrochloreWSM, Balents2013, YamajiImada2014, WitczakKrempa2014, Savary2014, GoswamiRoyDasSarma2017, Boettcher2017, GangChen2018, Ladovrechis2021} and Kondo-Weyl semimetals~\cite{Lai2017, Chen2022, Dzsaber2017}.

At the charge neutrality, the vanishing density of states (DOS) suppresses conventional Thomas-Fermi screening, rendering long-range Coulomb interactions a natural and often unavoidable ingredient in WSMs. In the simplest (linear) Weyl or Dirac systems, Coulomb interaction is marginal at tree level and flow to weak coupling, producing ``marginal Weyl/Dirac-liquid" with characteristic logarithmic corrections rather than a stable strong-coupling fixed point~\cite{Goswami2011, Hosur2012, Rosenstein2013, Isobe-Nagaosa-2013, DasSarma2015, roy2017_optical}. An especially rich extension is provided by generalized Weyl semimetals, where each node carries a higher integer monopole charge $\pm n$, protected by crystalline symmetries~\cite{Xu2011ChernSemimetalHgCr2Se4, Fang2012MultiWeyl, Huang2016, Zunger2017, Yang2014}, realizable in magnetic and noncentrosymmetric systems. The low-energy dispersion then becomes intrinsically anisotropic: linear along a momentum direction separating two Weyl nodes and nonlinear of order $n$ in the transverse plane. Such features trigger a sharp question.~\emph{How do long-range Coulomb interactions self-consistently reshape screening and quasiparticle coherence in generalized WSMs with arbitrary monopole charge $n$?} Search for an answer to this question propels the present quest.

Addressing this question requires theoretical control that simultaneously accounts for anisotropic scaling and gauge constraints. Because the Coulomb field is the temporal component of an underlying U(1) photon gauge field, the fermion self-energy and the fermion--boson vertex must satisfy the Ward-Takahashi identity \cite{Ward1950Identity, Takahashi1957GeneralizedWard}. A naive hard cutoff in all loop-momentum components typically violates such an identity by generating spurious scheme-dependent finite terms. We, therefore, formulate a Wilsonian renormalization group (RG) analysis tailored to the generalized Weyl dispersion, whose anisotropy singles out a hidden $(1+1)$-dimensional Lorentz-symmetric sector. We then fix a gauge-consistent UV regularization, adopted throughout this work, by benchmarking the Coulomb problem against the instantaneous analog of the Schwinger model~\cite{Schwinger1962GaugeInvarianceMassII}, in which one-loop polarization generates a finite photon mass that is uniquely fixed by gauge invariance in such a reduced $(1+1)$-dimensional theory.

In this Letter, we study the fate of quasiparticles in generalized Weyl semimetals with monopole charge $|n|>1$ in the presence of instantaneous long-range Coulomb interactions. We use a controlled large-$N$ expansion (with $N$ as the Weyl fermion flavors) in which screening is incorporated at the leading order via a dressed interaction within the random phase approximation (RPA), while the fermion self-energy and anomalous dimensions arise at ${\cal O}(1/N)$. We find an extended interaction-dominated scaling regime where the interplay between dispersion anisotropy and long-range interactions renders Coulomb screening intrinsically anisotropic, driving the effective interaction toward an emergent ``cylindrical'' form at long wavelengths. In this regime the fermions acquire a finite anomalous dimension and the quasiparticle residue is suppressed, yielding an \emph{anisotropic marginal Fermi liquid} (MFL) at intermediate energies. Although the Coulomb coupling is ultimately marginally irrelevant and flows to zero, it does so only logarithmically slowly, so the MFL  behavior manifests as a broad crossover controlled by a slowly running coupling rather than a stable strong-coupling fixed point. Such an interaction-driven crossover yields scaling in thermodynamics (specific heat and compressibility) and direction-dependent transport (optical conductivity) with multiplicative logarithmic corrections, resulting from the marginally irrelevant flow of the effective fine structure constant. Single-particle probes should likewise detect the loss of coherence via suppressed spectral weight and anisotropic linewidth broadening, measurable in angle-resolved photoemission spectroscopy (ARPES) over a parametrically broad intermediate-energy window.

A comment contrasting our findings with other works is due at this stage. Previously, the effects of the long-range Coulomb interaction have been studied by invoking RPA-dressed soft photon propagator~\cite{JianYao2015DoubleWeylCoulomb, Lai2015DoubleWeyl, ZhangJianYao2017TripleWeylCoulomb}, which, however, completely ignores its frequency dependence in an \emph{ad hoc} fashion, thereby neglecting any dynamic screening and forbidding fermions from acquiring any anomalous dimension. Hence, our conclusions are qualitatively distinct from the ones in Refs.~\cite{JianYao2015DoubleWeylCoulomb, Lai2015DoubleWeyl, ZhangJianYao2017TripleWeylCoulomb}. The retardation effects have also been ignored near the WSM-band insulator critical point in the presence of long-range Coulomb interaction~\cite{YangNatPhys2014}, while in a two-dimenional anisotropic semimetal, the underlying hidden one-dimensional Lorentz-symmetric brane on which soft photons acquire a gauge-invariant mass to fix the regularization scheme remained unnoticed so far~\cite{IsobePRL2016}.

\emph{Model}.~Since the long-range Coulomb interaction does not couple the Weyl points, we focus on the effective single-particle Hamiltonian for a generalized WSM close to one node with the monopole charge $n$, which reads as
\begin{align}~\label{eq:H}
H_{n}(\bm{k})= A_n k_{\perp}^n\Big[\sigma_1 \cos \!\big(n \phi_{k}\big)+\sigma_2 \sin \!\big(n \phi_{k}\big)\Big]+\sigma_3 v_z k_z,
\end{align}
where vector Pauli matrix ${\boldsymbol \sigma}$ acts on the orbital degrees of freedom, $k_{\perp}^2=k_x^2+k_y^2$ and $\phi_{k}=\tan^{-1}(k_y/k_x)$. The spectrum $E_{\bm{k}}=\pm \sqrt{A_n^2 k_\perp^{2n}+v_z^2 k_z^2}$ is anisotropic with nonlinear [linear] dispersion in the $(k_x,k_y)$ plane [along $k_z$], implying an enhanced low-energy DOS $\rho(E)\sim |E|^{2/n}$ with increasing $n$, which amplifies Coulomb-induced screening and self-energy renormalization. See also Sec.~1 of the Supplemental Material (SM) for details~\cite{SM}.

The imaginary-time action for instantaneous long-range Coulomb interaction is $S= \int d \tau\; d^3 x \; \mathcal{L}$ with
\allowdisplaybreaks[4]
\begin{align}\label{eq:action}
\mathcal{L}= 
\sum_{a=1}^{N} \psi_a^{\dagger}\Big[\big(\partial_\tau+i g \Phi\big)+H_n(\vec{k} \to -i {\boldsymbol \nabla})\Big] \psi_a
+ \frac{(\nabla \Phi)^2}{2},
\end{align}
where $\psi_a$ labels $N$ two-component fermionic flavors, $\Phi$ is the scalar potential mediating the interaction. Retardation effects are neglected, as the typical Fermi velocities are much smaller than the speed of light. The bare inverse fermionic and photon propagators are, respectively,
\begin{align}~\label{eq:bare_props}
G_0^{-1}(k)&= i k_0 + A_n |\bm{K}|^n\,\bm{\sigma}_\perp\!\cdot\!\hat{\bm{n}}_{k}+ \sigma_3 v_z k_z,\nonumber\\
{\rm and}\,\, D_0^{-1}(k) &= |\bm{K}|^2+c\, k_z^2
\end{align}
with frequency-momentum vector $k=(k_0,\bm{K},k_z)$, $\bm{K}=(k_x,k_y)$, $\bm{\sigma}_\perp=(\sigma_1,\sigma_2)$, and $\hat{\bm{n}}_{k}=(\cos n\phi_k,\sin n\phi_k)$. The parameter $c$ captures  anisotropic Coulomb propagator  generated by fermionic screening~\cite{sur2019, UryszekPRB2020, Gabriel2025}. See also Sec.~2 of SM for details~\cite{SM}.

\emph{Gauge-consistent RG approach}.~We implement a Wilsonian RG by gradually integrating out fast Fourier modes within the thin shell $\Lambda e^{-\ell}<E_{\bm{k}}<\Lambda$ with $\ell=\log(\Lambda/\omega)\ll1$, $\Lambda$ as the ultraviolet (UV) cutoff, and $\omega$ denoting the running scale. The generalized Weyl dispersion implies the anisotropic scaling $k_0\to k_0 e^{-\ell}, k_z\to k_z e^{-\ell}, k_\perp\to k_\perp e^{-\ell/n}$, so that $A_n k_\perp^n$ and $v_z k_z$ scale identically~\cite{roy2019, RoyFoster2018}. At the tree level the Coulomb coupling then is marginal, with  quantum corrections generating logarithmic flows that ultimately make this coupling marginally irrelevant for any $n$, as we show below.

Because $\Phi$ is the temporal component of a photon gauge field of the underlying electromagnetic theory, the fermion-boson vertex and the fermionic self-energy must satisfy the Ward-Takahashi identity~\cite{Ward1950Identity, Takahashi1957GeneralizedWard}. To fix a consistent prescription we benchmark our regularization against the  {Schwinger problem} with instantaneous Coulomb interaction such that in the $(1+1)$-dimensional reduction in the $(k_0,k_z)$ plane, the one-loop polarization yields a finite photon mass uniquely fixed by gauge invariance~\cite{Schwinger1962GaugeInvarianceMassII}. Requiring that our procedure reproduces this gauge-invariant result selects a unique ordering of loop integrations. Namely, one must integrate first over the frequency, $-\infty<k_0<\infty$, then over the  $k_z$ momentum component, $-\infty<k_z<\infty$, and only in the last step impose a UV cutoff on the remaining directions. The ambiguity in the integration procedure in the transversal ($k_x,k_y$) plane is then fixed by using the simple WSM ($n=1$) as the benchmark, where after integrating over the $k_0$ and $k_z$, imposing the in-plane cutoff in the rotationally symmetric manner automatically ensures the Ward-Takahashi identity, see SM, Sec.~3. Guided by this benchmark procedure, we then adopt the corresponding \emph{cylindrical} prescription in $(3+1)$ space-(imaginary-)time dimensions. First, we perform the loop integrals over $(k_0,k_z)$ and subsequently impose the UV cutoff only on the transverse momentum in a thin shell $k_\perp\in[\Lambda_\perp e^{-d\ell},\Lambda_\perp]$ (equivalently, on the corresponding energy scale $A_n k_\perp^n$). This choice eliminates U(1) gauge-symmetry violating power-law divergences, which preserves the Ward-Takahashi identity throughout the RG analysis and isolates the universal logarithmic terms that control RG flows of the couplings.

\emph{Large-$N$ expansion}.~With the UV regularization fixed, we organize the RG in a controlled large-$N$ expansion with $g^2\sim 1/N$ and keep the fine structure constant $\alpha_{_N}(\ell)\equiv {N g^2(\ell)}/{v_z(\ell)}$ fixed,   which renders the bosonic polarization $\mathcal{O}(N^0)$, while the fermion self-energy (and thus anomalous dimensions) is $\mathcal{O}(1/N)$.

At leading order in $1/N$, screening is incorporated through the RPA-dressed Coulomb propagator
\begin{equation}\label{eq:Ddressed}
D^{-1}(k)=|\bm{K}|^{2}+c\,k_{z}^{2}-N\,\Pi(k).
\end{equation}
For generalized Weyl nodes ($n>1$), the one-loop polarization (see Eq.~(7.15) in the SM~\cite{SM}) takes the form
$\Pi(k)=\Pi_{\perp}(k)\,|\bm K|^2+\Pi_{z}(k_0,k_z)$, 
with transversal and longitudinal components, respectively, given by
\begin{align}\label{eq:Pi_na_def}
&\Pi_{\perp}(k) =- \frac{n^2\alpha_{_N}}{24\pi^2 N}\left[\ln\left(\frac{\Lambda^{2/n}}{A_n^{{2}/{n}}\bs{Q}^2+b^{(2)}_n[k_0^2+(v_zk_z)^2]^{1/n}}\right)\right.\nonumber\\ 
&\left.-\left(b^{(3)}_n\frac{(v_z k_z)^2}{k_0^2+(v_zk_z)^2} - \ln2\right) \right],  \:\: \text{and} \:\: \nonumber\\
&\Pi_{z}(k_0,k_z)
=-\,b^{(1)}_n\frac{\alpha_{_N}}{N A_n^{2/n}}\;
\frac{(v_z k_z)^2}{[k_0^2+(v_zk_z)^2]^{\,1-1/n}}\,,
\end{align}
with $b^{(i)}_n\in\mathbb{R}$, where $i=1,2,3$. In the static limit $k_0=0$, the polarization is $\Pi_{z}(0,k_z)\propto \alpha_{_N}|k_z|^{2/n}/N$, so the dressed Coulomb propagator acquires an additional $|k_z|^{2/n}$ longitudinal piece from the fermionic bubble.

The resulting screened interaction feeds back into the fermion self-energy at $\mathcal{O}(1/N)$, according to
\begin{equation}\label{eq:Sigma_generic_clean}
\Sigma(k)= i k_{0}\,\Sigma_{0}(\ell)
+ A_{n}|\bm{K}|^{n}\,\bm{\sigma}_\perp \!\cdot\!\hat{\bm{n}}_{k}\,\Sigma_{12}(\ell)
+ \sigma_{3} v_{z} k_{z}\,\Sigma_{3}(\ell),
\end{equation}
thereby determining the scale dependence of the wave function renormalization and band parameters. Throughout, the
cylindrical prescription preserves the Ward-Takahashi identity, fixing the vertex correction by the frequency part
of the self-energy, $\delta g(\ell)=\Sigma_{0}(\ell)$, and excluding an independent vertex renormalization.

\emph{RG flows}.~After integrating out a thin Wilsonian shell and rescaling the fields, the
running parameters obey
\begin{align}~\label{eq:RG_eqs}
\dot A_n &= \big[\gamma_{12}(\ell)-\gamma_0(\ell)\big]\,A_n,\quad
\dot v_z = \big[\gamma_{3}(\ell)-\gamma_0(\ell)\big]\,v_z,\nonumber\\
\dot c &= \Big[\frac{2}{n}-2+\delta_{3}(\ell)-\delta_{12}(\ell)\Big]\,c,\:\: \text{and} \:\:
\dot g = -\frac{\delta_{12}(\ell)}{2}\,g,
\end{align}
where $\dot X\equiv dX/d\ell$. Notice that the operator $\sim c$ is irrelevant for any $n>1$, and as such can be omitted. The scale-dependent coefficients are obtained by extracting the logarithmic parts of the  self-energy and polarization components, namely, 
\begin{equation}\label{eq:gamma_delta_defs_precise}
\Sigma_a=\gamma_a(\ell)\,\ell\,\, (a=0,12,3)
\,\, \text{and} \,\,
\Pi_b(\ell)=\delta_b(\ell)\,{\ell}\,\,  (b=12,3),
\end{equation}
Here $\gamma_a(\ell)$ and $\delta_b(\ell)$ denote the coefficients multiplying the logarithmic contributions
produced by integrating out the infinitesimal Wilsonian shell, evaluated at fixed rescaled
external frequency and momenta. Within the large-$N$ expansion, the polarization is leading $\delta_b(\ell)=\mathcal{O}(\alpha_N)$, while the fermionic anomalous dimensions are subleading, $\gamma_a(\ell)=\mathcal{O}(\alpha_N/N)$, obtained from the one-loop self-energy evaluated with the dressed Coulomb propagator~\eqref{eq:Ddressed} (Table~\ref{tabe:CT} and Sec.~7 of the SM~\cite{SM}).

\begin{table}[t!]
\centering
\setlength{\tabcolsep}{3.5pt}
\renewcommand{\arraystretch}{1.15}
\small
\begin{tabular}{|c||c|c|c|c|}
\hline
CC & $n=1$ & $n=2$ & $n=3$ & $n=4$ \\
\hline\hline
$\gamma_0$  & $0$     & $0.344/N$ & $0.252/N$ & $0.165/N$ \\
\hline
$\gamma_{12}$ & $2.4/N$ & $0.747/N$ & $0.294/N$ & $0.263/N$ \\
\hline
$\gamma_{3}$  & $2.4/N$ & $1.172/N$ & $0.535/N$ & $0.298/N$ \\
\hline
$\delta_{12}$ & $\alpha_{_N}/12\pi^2$ & $2\alpha_{_N}/12\pi^2$ & $3\alpha_{_N}/12\pi^2$ & $4\alpha_{_N}/12\pi^2$ \\
\hline
$\delta_{3}$  & $v_z^2/(12\pi^2A_1^2)$ & $0$ & $0$ & $0$ \\
\hline
\end{tabular}
\caption{One-loop counterterm coefficients (CC) in Weyl semimetals with different integer monopole charge $n$. See also Secs.~3-7 of SM for details~\cite{SM}.}
\label{tabe:CT}
\end{table}

In the simple WSM ($n=1$), instantaneous Coulomb interaction produces an \emph{isotropic} MFL~\cite{Goswami2011, Hosur2012, Rosenstein2013, Isobe-Nagaosa-2013, DasSarma2015, roy2017_optical}. Within our cylindrical regularization one finds $\Sigma_0=0$ and hence $\gamma_0=0$, yielding $Z_\psi(\ell)=1$. The vertex correction vanishes since the Ward-Takahashi identity enforces $\delta g=\Sigma_0=0$. The remaining renormalizations dress the velocity isotropically, $\gamma_{12}=\gamma_3\propto \alpha_{_N}/N$, while the polarization generates isotropic logarithmic screening, with $\delta_{12}=\delta_3\propto \alpha_{_N}$. In contrast, for generalized WSMs ($n>1$) the qualitative changes, most notably $\gamma_0(\ell)>0$ and  anisotropic screening characterized by $\delta_{12}\sim\alpha_{_N}$ and  $\delta_3(\ell)=0$, as we show below, are intrinsic consequences of the higher monopole charge and the resulting anisotropic dispersion. For a generalized WSM, combining the RG equations for $g$ and $v_z$ yields the flow for $\alpha_{_N}$, given by
\begin{equation}\label{eq:alpha_flow}
\dot \alpha_N = -\big[\gamma_3-\gamma_0\big]\,\alpha_{_N}(\ell) - \delta_{12}(\ell)\,\alpha_{_N}(\ell).
\end{equation}
As a consequence, the long-range Coulomb interaction is \emph{marginally irrelevant}, implying that the effective coupling $\alpha_{_N}(\ell)$ flows to zero only logarithmically slowly as
$\ell\to\infty$ for any $n\ge2$ since  $ \delta_{12}>0$ and $\gamma_3>\gamma_0$, as shown in Table~\ref{tabe:CT}. Crucially, the slow (logarithmic) flow features a broad intermediate regime where $\alpha_{_N}(\ell)$ is finite and interaction effects therefore leave an imprint to the physical observables, as discussed below.

Remarkably, the Coulomb screening becomes sharply anisotropic as only the transverse sector is logarithmically dressed, according to 
\begin{equation}\label{eq:delta_n_short}
\delta_{12}(\ell)=\frac{n}{12\pi^2}\,\alpha_{_N}(\ell)
\,\, {\rm and}\,\,
\delta_3(\ell)=0,
\end{equation}
so that the fermion bubble generated part of the Coulomb interaction $\propto \alpha_{_N}|k_z|^{2/n}$ flows to smaller values as $\ell \to \infty$.  In this crossover regime, the static dressed Coulomb kernel is
$D_E^{-1}(\bm k)\;\simeq\; a_\perp(E)\,k_\perp^2 \;+\; a_z(E)\,|k_z|^{2/n}$, which in the  transverse plane $z=0$ leads to the  effective algebraic Coulomb tail, $V_E(r_\perp,0)\sim r_\perp^{-n}$ for $n>1$, before $\alpha_{_N}(\ell)\to0$  as $\ell\to\infty$. See Sec.~8 of SM for details~\cite{SM}.

\emph{Anisotropic MFL}.~A direct signature of the MFL is the progressive reduction of single-particle weight, quantified by the fermionic wavefunction renormalization $Z_\psi(\ell)$, given as $\gamma_0(\ell)=-d\ln Z_\Psi/d\ell$, which  inherits the scale dependence of the running Coulomb coupling  [see Eqs.~\eqref{eq:alpha_flow} and~\eqref{eq:gamma_delta_defs_precise}]. Notice that the  band parameters $A_n(\ell)$ and $v_z(\ell)$ are also renormalized, implying distinct interaction effects on the in-plane and out-of-plane fermion dynamics. Accordingly, the fermion Green's function takes  the RG-improved scaling form obtained by solving the corresponding Callan-Symanzik equation~\cite{peskin1995}  
\begin{equation}\label{eq:G_scaling}
G(\omega,\bm{k}) \sim
\frac{Z_\psi(\ell)}{i\omega - A_n(\ell)\, k_\perp^n\,\bm{\sigma}\!\cdot\!\hat{\bm{n}}_k - v_z(\ell)\, k_z \sigma_3},
\end{equation}
with $\ell \sim \ln\!\left(\frac{\Lambda}{\max\{|\omega|,\,E_{\bm{k}}\}}\right)$, making explicit that the single-particle response is controlled by the running parameters evaluated at the external scale. See also Sec.~9 of the SM for details~\cite{SM}. In the crossover regime, where $\alpha_{_N}(\ell)$ flows only logarithmically slowly, this produces an \emph{anisotropic} suppression of spectral weight and corresponding linewidth broadening. The intermediate-scale dynamics is then  characterized by values of the effective anomalous dimension (e.g., $\gamma_0\simeq 0.344/N$ for $n=2$ and $\gamma_0\simeq 0.252/N$ for $n=3$ in Table~\ref{tabe:CT}), which govern the suppression of the quasiparticle residue over a broad window of energy scales. As the infrared is approached, $\alpha_{_N}(\ell)\to 0$ and the system crosses over toward weak coupling, but this crossover can remain parametrically wide for moderate $N$ and realistic bare couplings.

\emph{Experimental implications}.~The marginal irrelevance of the Coulomb interaction implies that, at asymptotically low energies, the scaling behavior of observables is governed by free power laws dressed with universal logarithmic corrections. Evaluating running quantities at the external scale $E=\max\{T,\omega\}$ with $\ell_E\equiv \ln(\Lambda/E)$, the one-loop flow of the fine structure constant reads as
\begin{equation}~\label{eq:alpha_run_obs}
\alpha_{_N}(E)\simeq\frac{\alpha^0_{_N}}{1+\bar\delta_{12}\,
\alpha^0_{_N}\,\ln(\Lambda/E)}\,
\end{equation}
with $\delta_{12}={\bar\delta}_{12}\alpha_{_N}$ given by Eq.~\eqref{eq:delta_n_short}, so $\alpha_{_N}(E)\sim 1/\ln(\Lambda/E)$ at the lowest energies, with $\alpha^0_{_N} \equiv \alpha_{_N}(E=\Lambda)$ as the bare fine structure constant. In this weak-coupling regime $\gamma_i(\ell)\simeq \gamma_i\alpha_{_N}(\ell)=(\bar\gamma_i/N)\alpha_{_N}(\ell)$, and integrating the RG equations yields logarithmic renormalization at an energy scale $\omega$
\begin{equation}~\label{eq:logs_ZAv}
Z_\psi(\omega)\sim L_\omega^{-\eta_\psi},\,
A_n(\omega)\sim A_n\,L_\omega^{\eta_A},\, {\rm and}\,\,
v_z(\omega)\sim v_z\,L_\omega^{\eta_v},
\end{equation}
with
\begin{equation}\label{eq:etas_def}
\eta_\psi=\frac{\bar\gamma_0}{{\bar\delta}_{12}N},\,
\eta_A=\frac{\bar\gamma_{12}-\bar\gamma_0}{{\bar\delta}_{12} N},\, {\rm and}\,\,
\eta_v=\frac{\bar\gamma_{3}-\bar\gamma_0}{{\bar\delta}_{12} N},
\end{equation}
where $L_\omega=1+{\bar\delta}_{12}\alpha^0_{N}\ell_\omega$. These logarithmic corrections feed directly into observables through the renormalized DOS $\rho(\omega)\sim Z_\psi(\omega)\,\omega^{2/n}/\!\big[v_z(\omega)\,A_n(\omega)^{2/n}\big]$. Consequently, the specific heat and compressibility acquire explicit logarithmic corrections at temperature $T$
\begin{equation}~\label{eq:thermo_logs}
C(T)\sim \frac{T^{1+2/n}}{\left[\ln(\Lambda/T)\right]^{p_{\rm th}}}
\;\; {\rm and} \;\; 
\kappa(T) \sim \frac{T^{2/n}}{\left[ \ln(\Lambda/T) \right]^{p_{\rm th}}},
\end{equation}
respectively, where $p_{\rm th}=\eta_\psi+\eta_v+\frac{2}{n}\eta_A$.

Optical conductivity is anisotropic already at the bare level as can be inferred directly  using the interband Kubo formula   at $\mu=0$  (see Sec.~9 of the SM~\cite{SM}), leading to $\sigma_{xx}(\omega)\sim {\omega}/{v_z}$ and $\sigma_{zz}(\omega)\sim {v_z}{A_n^{-2/n}}\;\omega^{\frac{2}{n}-1}$~\cite{roy2019}. 
The marginally irrelevant Coulomb interaction produces multiplicative logarithmic corrections through the
running parameters at scale $\omega$, with the explicit form  
\begin{equation}~\label{eq:opt_logs_correct}
\sigma_{xx}(\omega) \sim \frac{\omega}{\left[\ln(\Lambda/\omega)\right]^{p_x}}
\:\:\: \text{and} \:\:\:
\sigma_{zz}(\omega) \sim \frac{\omega^{\frac{2}{n}-1}}{\left[ \ln(\Lambda/\omega) \right]^{p_z}},
\end{equation}
obtained after invoking Eq.~\eqref{eq:logs_ZAv}, with the corresponding exponents $p_x=2\eta_\psi+\eta_v$ and
$p_z=2\eta_\psi-\eta_v+\frac{2}{n}\eta_A$.

{Optical shear viscosity is also anisotropic already at the bare level, with the (non-vanishing) purely in-plane components at $\mu=0$ scaling as  $\eta_\perp\sim A_n^{-2/n}v_z^{-1}\omega^{1+2/n}$. For $n>1$, the components involving the $z$ direction split into two additional classes:
$\eta_{jzjz}\sim v_z^{-3}\omega^{3}$, and
$\eta_{zjzj}\sim v_z A_n^{-4/n}\omega^{4/n-1}$, with $j=x,y$, while $ \eta_{zzzz}\sim \eta_{jjzz}\sim \eta_{jzzj}\sim \eta_\perp\sim A_n^{-2/n}v_z^{-1}\omega^{1+2/n}$~\cite{moore-PRB-2020}.
Including the marginally irrelevant Coulomb interaction then yields the corrections to scaling  (see Sec.~9 of the SM~\cite{SM})
\begin{align}
&\eta_{\perp}(\omega)\sim \eta_{zzzz}\sim \eta_{jjzz}\sim \eta_{jzzj}
\sim \frac{\omega^{1+\frac{2}{n}}}{\left[\ln(\Lambda/\omega)\right]^{p_1}},\\
&\eta_{jzjz}(\omega)\sim \frac{\omega^{3}}{\left[\ln(\Lambda/\omega)\right]^{p_2}}\,\,\,{\rm and}\,\,\,
\eta_{zjzj}(\omega)\sim \frac{\omega^{\frac{4}{n}-1}}{\left[\ln(\Lambda/\omega)\right]^{p_3}}, \nonumber
\end{align}
with
$p_1=p_x+\frac{2}{n}\eta_A$, 
$p_2=p_x+2\eta_v$, 
$p_3=p_z+\frac{2}{n}\eta_A$, and exponents $p_{x}$ and $p_z$ defined in Eq.~\eqref{eq:opt_logs_correct}.}

Finally, single-particle probes provide a direct handle on the MFL. In particular, ARPES measures a quasiparticle residue suppressed by the marginally irrelevant Coulomb interaction, $Z_\psi(\omega)\sim[\ln(\Lambda/\omega)]^{-\eta_\psi}$, accompanied by anisotropic linewidth broadening. Although the asymptotic suppression is only logarithmic, over the experimentally accessible range the slow RG flow of $\alpha_{_N}(E)$ makes the logarithmic corrections vary weakly, so the suppression of the spectral weight  can be well fit by a power law with a slowly running  effective exponent $\eta_{\psi,\rm eff}(E)\sim 1/\ln(\Lambda/E)$, see Sec.~9.6 of the SM~\cite{SM}.

\emph{Conclusions and discussion}.~Here, we analyze generalized Weyl semimetals with instantaneous long-range Coulomb interactions using a gauge-consistent Wilsonian RG scheme that preserves the Ward-Takahashi identity. Within a controlled large-$N$ expansion, we uncover a qualitative distinction between simple ($n=1$) and higher-charge ($n>1$) Weyl nodes. Whereas the $n=1$ case retains isotropic (marginal) Weyl-liquid character, the $n>1$ systems develop an interaction-dominated anisotropic scaling regime with a finite fermionic anomalous dimension and a pronounced suppression of the quasiparticle residue, yielding \emph{an anisotropic MFL} over a broad intermediate-energy window.

This outcome is a consequence of the emergence of \emph{intrinsically anisotropic screening} for $n>1$. Namely, fermionic polarization produces logarithmic dressing in the transverse sector but not in the longitudinal one as $\delta_{12}(\ell) \neq 0$ and $\delta_{3}(\ell)=0$, driving the Coulomb propagator toward an effectively anisotropic long-wavelength form. The Coulomb coupling is nevertheless marginally irrelevant and flows to zero only logarithmically slowly, which generates a parametrically wide crossover window where interaction effects remain sizable, implying that multiplicative logarithmic corrections can mimic effective power laws over finite scales.

Experimentally, the anisotropic MLF should manifest as direction-dependent spectral-weight suppression and linewidth broadening in ARPES, and through the scaling of transport responses, dressed by logarithmic corrections. Finally, our results should hold even when the chemical potential ($\mu$) is placed slightly away from the Weyl nodes, up to an infrared scale set by $\mu$. 

\emph{Acknowledgment}.~This work was supported by Fondecyt (Chile) Grant No.~1230933 (V.J.) and Grant No.~1241033 (R.S.-G.). B.R.\ was supported by NSF CAREER Grant No.\ DMR-2238679 and thanks ANRF, India, for support through the Vajra scheme VJR/2022/000022.

\emph{Data availability.} The data that support the findings of this
article are available from the authors upon reasonable request.

\bibliography{ref}

\end{document}